\documentclass[manuscript]{aastex}

\newcommand{\neii}{[Ne~{\sc ii}]}
\newcommand{\uflux}{erg\,cm$^{-2}$\,s$^{-1}$}  

\slugcomment{$^\star$Based on observations made with VISIR on the UT3/Melipal ESO Telescope at Paranal under programme ID 080.C-0404(A)}

\shorttitle{Central star-driven photoevporation}
\shortauthors{Pascucci \& Sterzik}

\begin{document}

%% LaTeX will automatically break titles if they run longer than
%% one line. However, you may use \\ to force a line break if
%% you desire.

\title{Evidence for Disk Photoevaporation Driven by the Central Star$^\star$}

%% Use \author, \affil, and the \and command to format
%% author and affiliation information.
%% Note that \email has replaced the old \authoremail command
%% from AASTeX v4.0. You can use \email to mark an email address
%% anywhere in the paper, not just in the front matter.
%% As in the title, use \\ to force line breaks.

\author{I. Pascucci}
\affil{Department of Physics and Astronomy, Johns Hopkins University, Baltimore, MD 21218}
\email{pascucci@pha.jhu.edu}

\and

\author{M. Sterzik}
\affil{European Southern Observatory, Casilla 19001, Santiago 19, Chile }
\email{msterzik@eso.org}

%% Notice that each of these authors has alternate affiliations, which
%% are identified by the \altaffilmark after each name.  Specify alternate
%% affiliation information with \altaffiltext, with one command per each
%% affiliation.

%\altaffiltext{1}{Visiting Astronomer, Cerro Tololo Inter-American Observatory.
%CTIO is operated by AURA, Inc.\ under contract to the National Science
%Foundation.}

%% Mark off your abstract in the ``abstract'' environment. In the manuscript
%% style, abstract will output a Received/Accepted line after the
%% title and affiliation information. No date will appear since the author
%% does not have this information. The dates will be filled in by the
%% editorial office after submission.

\begin{abstract}
The lifetime of isolated protoplanetary disks is thought to be set by the combination of viscous accretion and photoevaporation driven by stellar high-energy photons. Observational evidence for magnetospheric accretion in young sun-like stars is robust. Here we report the first observational evidence for disk photoevaporation driven by the central star. 
We acquired high-resolution (R$\sim$30,000) spectra of the \neii{} 12.81\micron{} line from 7 circumstellar disks using VISIR on Melipal/VLT. We show that the 3 transition disks in the sample all have \neii{} line profiles consistent with those predicted by a photoevaporative flow driven by stellar  extreme UV photons. The $\sim$6\,km/s blue-shift of the line from the almost face-on disk of TW~Hya is clearly inconsistent with emission from a static disk atmosphere and convincingly points  to the presence of a photoevaporative wind. We do not detect any \neii{} line close to the stellar velocity from the sample of classical optically thick (non-transition) disks. 
% have \neii{} lines near the star velocity, in spite of spectrally unresolved \neii{} fluxes comparable to those of transition disks. 
We conclude that most of the spectrally unresolved \neii{} emission in these less evolved systems arises from jets/outflows rather than from the disk. The pattern of the \neii{} detections and non-detections suggests that extreme UV-driven photoevaporation starts only at a later stage in the disk evolution.
\end{abstract}

%% Keywords should appear after the \end{abstract} command. The uncommented
%% example has been keyed in ApJ style. See the instructions to authors
%% for the journal to which you are submitting your paper to determine
%% what keyword punctuation is appropriate.


%% From the front matter, we move on to the body of the paper.
%% In the first two sections, notice the use of the natbib \citep
%% and \citet commands to identify citations.  The citations are
%% tied to the reference list via symbolic KEYs. The KEY corresponds
%% to the KEY in the \bibitem in the reference list below. We have
%% chosen the first three characters of the first author's name plus
%% the last two numeral of the year of publication as our KEY for
%% each reference.

%% Authors who wish to have the most important objects in their paper
%% linked in the electronic edition to a data center may do so by tagging
%% their objects with \objectname{} or \object{}.  Each macro takes the
%% object name as its required argument. The optional, square-bracket 
%% argument should be used in cases where the data center identification
%% differs from what is to be printed in the paper.  The text appearing 
%% in curly braces is what will appear in print in the published paper. 
%% If the object name is recognized by the data centers, it will be linked
%% in the electronic edition to the object data available at the data centers  
%%
%% Note that for sources with brackets in their names, e.g. [WEG2004] 14h-090,
%% the brackets must be escaped with backslashes when used in the first
%% square-bracket argument, for instance, \object[\[WEG2004\] 14h-090]{90}).
%%  Otherwise, LaTeX will issue an error. 

\section{Introduction}
It is well established that most $\sim$1 Myr-old stars are surrounded by relatively massive optically thick dust disks. By an age of $\sim$10\,Myr only a few percent of sun-like stars still retain an optically thick dust disk (e.g. \citealt{hernandez07}) and many intermediate-mass stars might already harbor second-generation dust disks (e.g. \citealt{currie08}). This fast clearing of primordial dust is in agreement with the short formation timescales of chondrules, asteroids, and planets in the Solar System \citep{pascucci09}.
Although much less is known about the evolution of the gas component, there are at least three observables pointing to a similarly fast dispersal timescale: a) an order of magnitude lower accretion rate for the few stars still accreting at an age of $\sim$10\,Myr in comparison to $\sim$1\,Myr-old stars \citep{muzerolle00,lawson04}; b) gas mass upper limits of less than 0.1\,M$_{\rm Jup}$ in disks around non-accreting stars \citep{holl05,pascucci06}; and c) upper limits on the H$_2$-to-dust ratio of less than 10 in two $\sim$\,12\,Myr-old edge-on disks \citep{lecavelier01,roberge05}. Which are the physical mechanisms clearing out primordial disks?

Models of protoplanetary disk evolution suggest that most of the disk mass is cleared out by viscous evolution (accretion of gas onto the central star) and photoevaporation driven by the central star (e.g., \citealt{clarke01,alexander06b,gorti09}). %Indeed, models that combine these dispersal mechanisms can qualitatively reproduce two well-established observational trends (e.g. Dullemond et al. ): (a) a rapid disk dispersal compared to the disk lifetime; and (b) the almost simultaneous loss of the inner and the outer disks. 
The strongest evidence that young stars are accreting nebular gas comes from the veiling of optical and UV photospheric absorption lines which is produced by the continuum emission of an accretion shock at the stellar surface (e.g. \citealt{calvet00}). In addition, the large (few hundred km/s) line widths and asymmetries of permitted emission lines are found to be well reproduced by models with infalling gas via magnetospheric accretion \citep{muzerolle98}. In contrast to the numerous diagnostics of accretion, unambiguous diagnostics of centrally-driven disk photoevaporation are lacking. 

Here we show that the profiles of \neii{} forbidden lines at 12.81\,\micron{} from disks with dust inner holes (hereafter, transition disks) are entirely consistent with those predicted from photoevaporative disk winds (Sect.~\ref{Sect:transition}). These observations provide the first strong evidence in favor of centrally-driven disk photoevaporation. 
We also demonstrate that in disks with radially continuous optically thick dust (classical, non-transition disks) most of the \neii{} emission detected with the Spitzer Space Telescope does not arise in a disk but likely in a jet/outflow (Sect.~\ref{Sect:thick}). Finally, we discuss how observations, such as those presented in this paper, can be used to identify the onset of disk photoevaporation (Sect.~\ref{Sect:discussion}).

\section{Motivation, Observations, and Data Reduction}\label{Sect:Observations}
The IRS spectrograph on board the Spitzer Space Telescope has recently
detected the \neii{} 12.81\,\micron{} line toward many young protoplanetary
disks as well as toward the possibly more evolved transition disks 
\citep{geers06,pascucci07,lahuis07,espaillat07,ratzka07,flaccomio09,guedel09}.
Lines appear spatially and spectrally unresolved indicating that the
\neii{} emission is confined to within $\sim$1,000\,AU from the central
star and lines are narrower than $\sim$500\,km/s. Models of disks
irradiated by stellar X-rays or extreme UV (EUV, $h \nu >$ 13.6\,eV)
photons can sufficiently ionize the disk surface to reproduce the observed
line fluxes \citep{glassgold07,meijerink08,gorti08,ercolano08,hollenbach09} suggesting that \neii{} emission is a robust tracer of disk gas. However, the first two spectrally resolved \neii{} lines brought some surprises. In the case of TW~Hya, \citet{herczeg07} found that the \neii{} line is centered at the stellar radial velocity, consistent with a disk origin, but about a factor of two broader than that predicted from X-ray and EUV irradiated disks. \citet{van09} spatially and spectrally resolved the strong \neii{} line from the T Tau complex demonstrating that in this system most of the \neii{} emission originates in jets/outflows with neon atoms partly ionized by stellar X-rays. 

Our aim here is to enlarge the sample of spectrally resolved \neii{} lines and understand in which systems \neii{} emission originates in a disk.  In the following subsections we describe our observational campaign and the data reduction (Sects.~\ref{sect:targets},~\ref{sect:observations},~\ref{sect:dataredu}). Observations were carried out with the high-resolution (R$\sim$30,000) spectrograph VISIR mounted on the VLT telescope {\it Melipal} (Sect.~\ref{sect:observations} for more details). In addition to the already published Spitzer/IRS spectra, we reduced and present here new archival IRS spectra that aid the interpretation of the VISIR observations (Sect.~\ref{sect:dataredu}).

\subsection{Target Selection}\label{sect:targets}
Because of the lower-sensitivity of VISIR with respect to Spitzer/IRS, we restricted our sample to
disks with bright/unresolved \neii{} lines (Spitzer fluxes greater than 10$^{-14}$\,\uflux). In
addition, we selected disks brighter than $\sim$50\,mJy in the continuum because target acquisition for
fainter objects becomes challenging with VISIR. The 6 targets we selected for this campaign are TW~Hya,
CS~Cha, VW~Cha, T~Cha, Sz~73, and Sz~102 (see Table~\ref{table:prop} and ~\ref{table:modelparameters}
for their main properties). At the beginning of the second night we also observed the double-lined
spectroscopic binary HD~34700A, which is the brightest infrared T~Tauri couple in a quadruple system
\citep{sterzik05}. Although HD~34700A did not have a published \neii{} line detection, its near- and
mid-infrared spectra present unusually strong PAH emission bands (e.g., \citealt{smith04}), possibly
hinting to a high stellar far-UV flux (e.g., \citealt{geers06}), which may be indicative of a high EUV flux. 
%We reduced the low-resolution Spitzer spectrum available in the archive following the procedure outlined in xx and zz. We found no hint of \neii{} emission line {\bf maybe need to reduce the high-res??} but we used the spectrum to flux calibrate the VISIR data (see below).

%%%%%%%% Sample properties
The systems we selected can be grouped into three categories based on their broad-band spectral energy distribution (SED). VW~Cha, Sz~73, and Sz~102 have significant excess emission relative to the photospheric flux from near- through to far-infrared wavelengths  \citep{gauvin92,hughes94,kessler06}. Such broad SEDs can be well reproduced by {\it radially continuous optically thick dust disks}, extending from  a few stellar radii out to hundreds of AU. TW~Hya, CS~Cha, and T~Cha are classified as {\it transition disks} in the literature. Their SEDs present a strongly reduced (or lack of) near-infrared excess emission but large mid- and far-infrared emission. Detailed modeling of their SEDs points to relatively large inner dust cavities almost devoid of sub-micron- and micron-sized dust grains: $\sim$1-4\,AU for TW~Hya \citep{calvet02,ratzka07}, $\sim$43\,AU for CS~Cha \citep{espaillat07}, and $\sim$15\,AU for T~Cha \citep{brown07}. There is no evidence of gas inner holes in these systems: TW~Hya is accreting disk gas at a rate of $\sim 5 \times 10^{-10}$ M$_{\sun}$/yr \citep{muzerolle00}, the spectroscopic binary CS~Cha at $< 10^{-8}$ M$_{\sun}$/yr \citep{espaillat07}\footnote{the estimate from \citet{espaillat07} should be considered only an upper limit because the UV excess emission was computed assuming that CS~Cha is a single star}, while the small ($< 10$\AA) but variable H$\alpha$ equivalent width from T~Cha suggests low-level and possibly episodic accretion \citep{alcala93}.  Finally, the SED of HD~34700A has little excess emission at all wavelengths produced by a tenuous dust disk, possibly a {\it debris disk} \citep{sylvester96}.

\subsection{Observations}\label{sect:observations}
We performed long-slit high-resolution spectroscopy with the spectrograph VISIR mounted on the VLT telescope {\it Melipal}  \citep{lagage04}. The observations were executed on 20, 21, and 22 February 2008. VISIR has a $256\times256$ BIB detector array with an image scale of 0\farcs127/pixel. We acquired all targets with the PAH2-NEII filter. We used a  slit width of 0\farcs4 centered at 12.81\micron{} to detect and spectrally resolve the \neii{} ($^2P_{3/2}-^2P_{1/2}$) fine-structure line at 12.81355\micron{} \citep{yamada85}. With this configuration we covered the spectral region between $\sim12.79$ and 12.83\,\micron{} with a resolution of $R\sim30,000$ as measured from first light data \citep{kaufl06} and from the FWHM of 5 narrow O$_3$ sky lines from our spectra. This resolution corresponds to 3 VISIR pixels or $\sim$10\,km/s in velocity scale.
 
We applied the standard chopping/nodding technique to suppress the mid-infrared background. For each target the slit was positioned in the default North-South orientation and chopping/nodding was done along the slit with a throw of 8\arcsec . Total on-source exposure times per science target were 1\,h or longer. To correct the spectra for telluric absorption and to obtain an absolute flux calibration, we observed standard stars immediately before or after the science target and at a similar airmass. A summary of the observations is presented in Table~\ref{table:log}.

\subsection{Data Reduction}\label{sect:dataredu}
{\bf VISIR high-resolution spectra.} For each nod position VISIR raw data are organized in cubes with the third axis having $2n+1$ planes, where $n$ is the number of chopping cycles. 
%(in our case 2). 
For each chopping cycle ($i$) two so-called Half-Cycle exposures are made: the $A_i$ image from the on-source position of the chopper, and the $B_i$ image from the off-source position of the chopper. The data cube stores the $A_i$ images in the odd planes and the average of the current and all previous $A_i-B_i$ images in the even planes. The last plane of the cube contains the average of all $A_i-B_i$ images and it is thus identical to the $2 \times n$ plane.

We process the raw data cubes with the VISIR pipeline version 3.2.1 \citep{lundin08}. First, 2D frames in each nod position are created by averaging all the Half-Cycle difference images in the data cube and low-level vertical stripes are removed. This step results in a set of images with a positive and a negative beam because the chop throw we chose is smaller than half of the VISIR field-of-view. Then, nodded images (images corrected for the telescope nodding) are produced by averaging the images in the two nod positions (A and B) and dividing them by 2 times the detector integration time. At this point each nodded image contains a double positive beam in the center and a negative beam on each side separated by the nodding(=chopping) throw. Pixels detected as bad in the Half-Cycle frames are now cleaned by interpolation with neighboring pixels. 
%% bad pixels remain stable in these images they are cleaned now because other pixels are more stable at this point
In parallel, a reference frame of the infrared background, not corrected for chopping or nodding, is also created from the first Half-Cycle of the first nod position.  
%% There are many half-cycle images but we don't know if they are all used for the wavelength calibration 
The nodded images as well as the infrared background image are corrected for the optical distortion, which is known analytically, and are shifted and added to form the final combined image using the offsets stored in the FITS header (the nodding cycle sequence used for VISR is AnBnBnAn). The last step before extraction is to fold the two negative beams into the central positive beam. 

\begin{figure}
\includegraphics[angle=90,scale=.50]{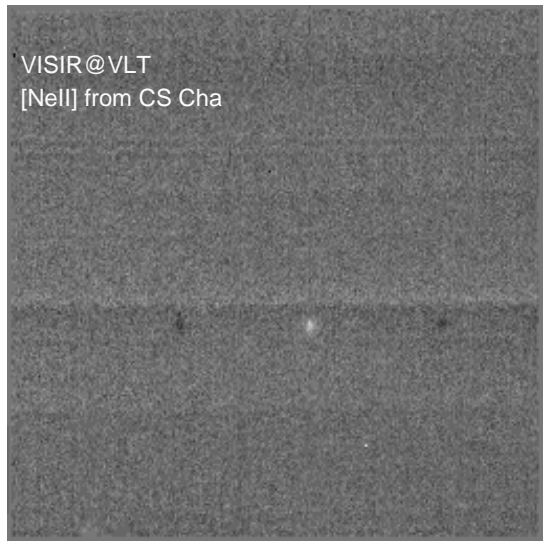}
\caption{Combined image of CS~Cha from the first dataset (February 22). The \neii{} emission is clearly detected both in the 
positive and in the negative beams but there is no detection of the continuum emission. The combined images from the other two datasets are similar and therefore not shown here. \label{figure:cscha}
}
\end{figure}

The spectrum is then extracted following the optimal extraction method by \citet{horne86}. In brief, the image is collapsed in the spectral dispersion direction to create a source profile, which is normalized and then expanded in spectral direction to obtain a weight map. The science frame is multiplied by the weight map and the spectrum is obtained by summing in spatial direction. This extraction method works well for sources where, in addition to narrow emission line(s), the continuum emission is also detected. For sources like CS~Cha, where only a narrow emission line is detected (but not the continuum, see Fig.~\ref{figure:cscha}), the optimal extraction method fails. For this source and its calibrators we have performed the standard full aperture extraction (same weights within a defined spatial range of pixels). We tested that an aperture of 10 pixels gives the best S/N in the extracted spectra.  

The wavelength calibration is done using the collapsed background frame in spatial direction and cross-correlating it to a synthetic model spectrum of the atmosphere. The dispersion relation is well approximated by a first order polynomial and the offset in pixels that maximizes the cross-correlation can be determined to an accuracy of 0.01 pixel ($\sim$0.03\,km/s). 
We also compared the observed peak position of two photospheric absorption
lines in the spectra of the K5 III star HD 136422 and the K4.5 III star HD
139127 to those from the MARCS model atmosphere of a K5 III star (HR 6705,
the spectrum was kindly provided by L. Decin). This exercise demonstrates
that the peak centroids can be determined to a lower accuracy, varying
from a tenth up to a few km/s (see also Table 3).
%We have tested that this accuracy applies to our dataset. To do that we
%compared the observed peak position of 2 photospheric absorption lines in
%the spectra of the K5 III star HD~136422 to those from the MARCS model
%atmosphere of a K5 III star (HR~6705, the spectrum was kindly provided by
%L. Decin). The peak centroids in the stellocentric frame are shifted by at
%most 0.034\,km/s, confirming the wavelength accuracy from the
%cross-correlation technique.  

All steps described above have been applied to both the target and its standard star(s). In addition, the extracted spectrum of the standard star was flux calibrated using the model flux from the VISIR standard star catalog. From this calibration we created a spectral response function (Jy/ADU) that we applied at the wavelengths of the science spectrum. This step removes most of the ``fringing'' present in the VISIR spectra (the flux modulation with wavelengths has a peak to peak amplitude of about 15\% and is found to be stable over periods of at least several months, \citealt{van09}). %We have tested on the spectrum of TW~Hya a different calibration procedure which corrects for telluric lines with ATRAN model atmospheres (that give the best match to calibration observations) and flux calibrates the science spectra with standard stars. In this approach we had to correct for the fringing pattern. To characterize the fringing pattern we fitted a spline profile to the two high S/N spectra from the calibrator HD~136422. We have then divided the science and calibrator observations by this profile. Tests on TWHya show that spectra calibrated with these two different methods agree well within the estimated errorbars. This second approach was used for the targets HD~34700, T~Cha, Sz~102, and Sz~73 because their calibrators have spectra with either low S/N  or photospheric absorption features. 

{\bf Spitzer/IRS spectra.} As mentioned in Sect.~\ref{sect:targets}, all
targets except HD~34700 have published low- (R$\sim$105) or
high-resolution (R$\sim$600) IRS spectra with bright \neii{} emission
lines (see Table~\ref{table:prop}). We have reduced the archival IRS
low-resolution spectrum of HD~34700 and report no detection of the \neii{}
line. In addition, we have reduced archival high-resolution IRS spectra
for TW~Hya and CS~Cha obtained at different epochs than the already
published spectra by \citet{ratzka07} and \citet{espaillat07}. These
spectra were acquired as part of the Spitzer GO program 30300 (PI, J.
Najita) and are presented here to aid the interpretation of the VISIR
results (Sect.~\ref{Sect:transition}). The data reduction of the IRS high-
and low-resolution spectra followed the procedure outlined in
\citet{pascucci06,pascucci07,pascucci08} and \citet{bouwman08}. The
\neii{} line fluxes from these new spectra are also reported in
Table~\ref{table:prop} (first entries in column 5).

%% In a manner similar to \objectname authors can provide links to dataset
%% hosted at participating data centers via the \dataset{} command.  The
%% second curly bracket argument is printed in the text while the first
%% parentheses argument serves as the valid data set identifier.  Large
%% lists of data set are best provided in a table (see Table 3 for an example).
%% Valid data set identifiers should be obtained from the data center that
%% is currently hosting the data.
%%
%% Note that AASTeX interprets everything between the curly braces in the 
%% macro as regular text, so any special characters, e.g. "#" or "_," must be 
%% preceded by a backslash. Otherwise, you will get a LaTeX error when you 
%% compile your manuscript.  Special characters do not 
%% need to be escaped in the optional, square-bracket argument.

%% In this section, we use  the \subsection command to set off
%% a subsection.  \footnote is used to insert a footnote to the text.

%% Observe the use of the LaTeX \label
%% command after the \subsection to give a symbolic KEY to the
%% subsection for cross-referencing in a \ref command.
%% You can use LaTeX's \ref and \label commands to keep track of
%% cross-references to sections, equations, tables, and figures.
%% That way, if you change the order of any elements, LaTeX will
%% automatically renumber them.

%% This section also includes several of the displayed math environments
%% mentioned in the Author Guide.

\section{Results}\label{sect:results}
\begin{figure}
\includegraphics[scale=0.8]{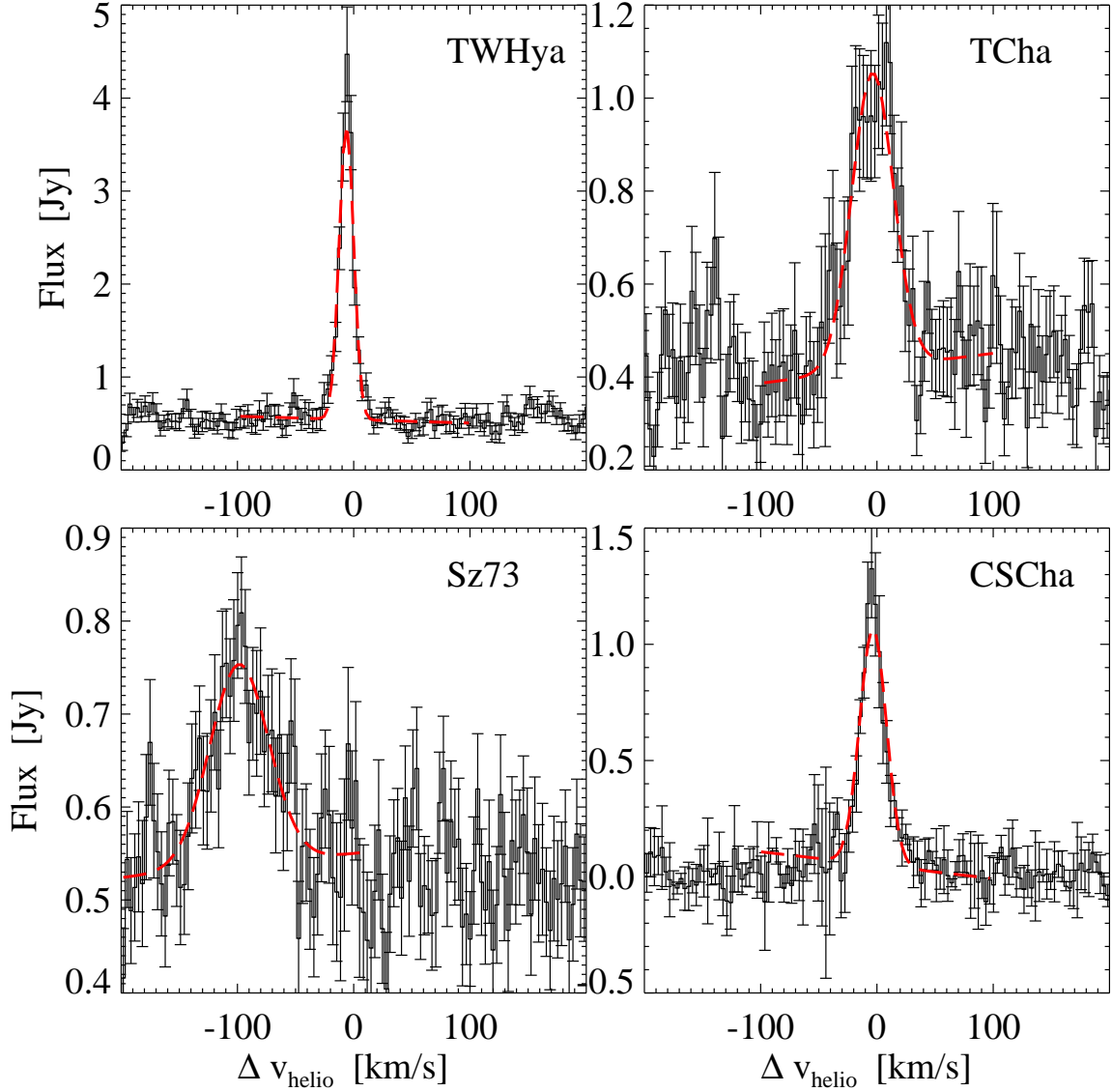}
\caption{Sources for which we detect \neii{} emission lines in their VISIR spectra. The x-axis gives the velocity in the stellocentric frame (see Table~\ref{table:prop} for the stellar heliocentric radial velocity). On top of the continuum emission we overplot the
best Gaussian fits to the data (dashed lines). In the case of CS~Cha the continuum emission is not detected (see also Fig.~\ref{figure:cscha}). \label{figure:det}
}
\end{figure}

\begin{figure}
\includegraphics[scale=0.8]{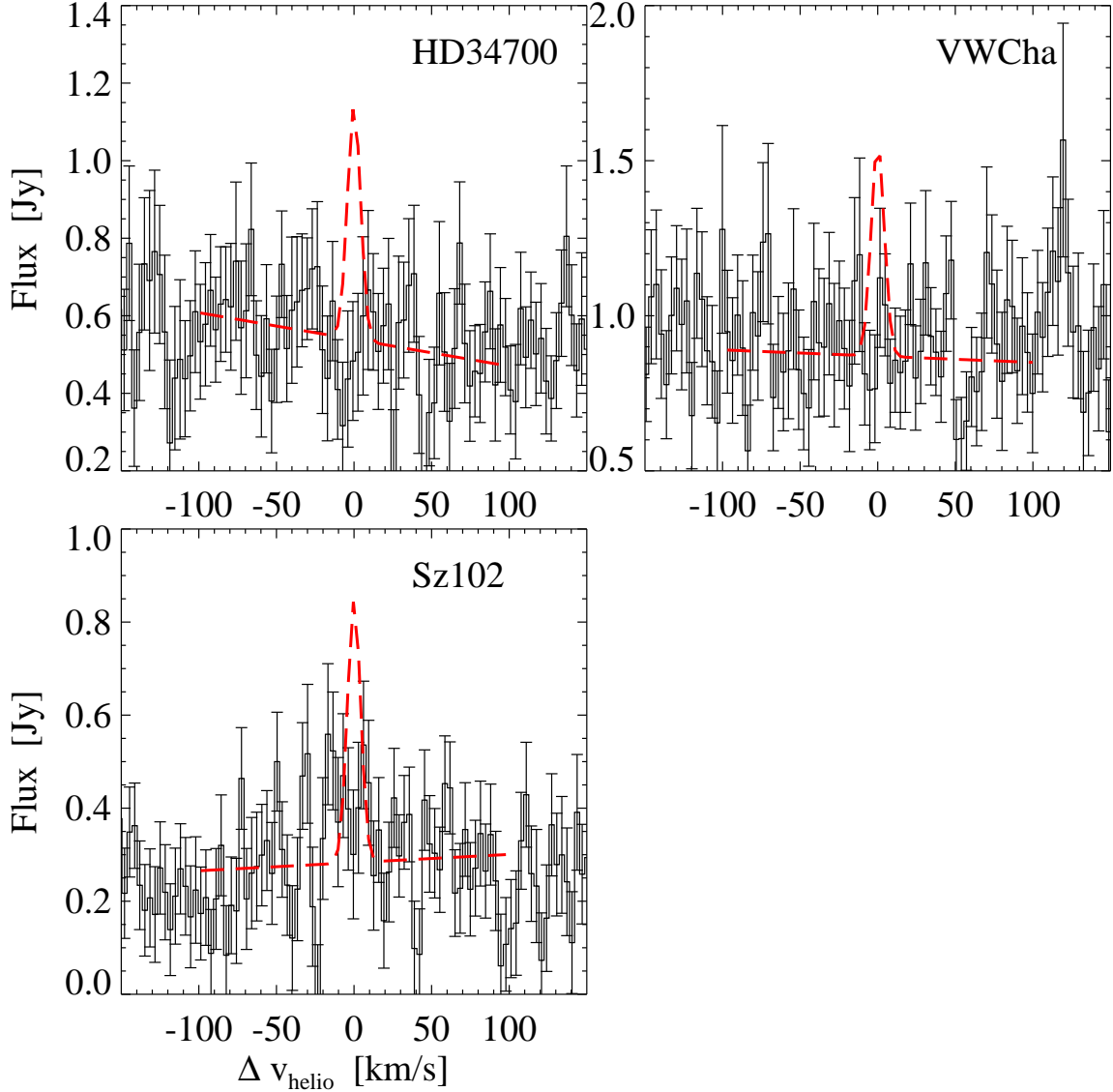}
\caption{Sources for which we do not detect any \neii{} emission line in their VISIR spectra. The x-axis gives the velocity in the stellocentric frame. On top of the continuum emission we overplot the hypothetical 3$\sigma$ upper limits (dashed lines) reported in Table~\ref{table:results} for line widths equal to 10\,km/s. 
Note that spectrally unresolved \neii{} lines have been detected with Spitzer toward  VW~Cha and Sz~102 (Table~\ref{table:prop}).
Our non-detections suggest that most of the \neii{} emission measured with the Spitzer Space Telescope is not originating in a disk but likely in a jet/ouflow. \label{figure:Ndet}
}
\end{figure}

We detect and spectrally resolve \neii{} emission lines from 4 out of 7 targets, specifically  TW~Hya, CS~Cha, T~Cha, and  Sz~73. 
%To make a more direct comparison between our and the Spitzer results, we scale the VISIR continuum to that measured in the Spitzer spectra (Sect.~\ref{sect:dataredu} for HD~34700, TW~Hya, and CS~Cha and \citealt{lahuis07} for  Sz~73, Sz~102, T Cha, and VW~Cha).
The \neii{} emission detected with VISIR is found to be comparable to the angular resolution  ($\sim$0.5\arcsec) estimated from both the mid-infrared continuum of our targets and from the telluric standards observed before or after the targets. 
%% Need to make a detailed study! For now I did it only for TWHya 
The profiles of the lines are consistent with a single Gaussian profile (see Fig.~\ref{figure:det}). Table~\ref{table:results} provides the peak centroids (in the stellocentric frame), FWHMs, and fluxes of the \neii{} lines computed assuming a Gaussian profile and a first-order polynomial for the continuum.
% Levenberg-Marquardt algorithm. 
In the case of non-detections (see Fig.~\ref{figure:Ndet}), we fit a first-order polynomial within $\pm$100\,km/s of the star velocity. Table~\ref{table:results} provides the 3$\sigma$ upper limits to the \neii{} flux computed from the RMS in the baseline-subtracted spectrum and assuming a line width of 10\,km/s, equal to the instrument resolution (Sect.~\ref{sect:observations}). Upper limits for broader lines simply scale by (FWHM/10). Figs.~\ref{figure:det} and~\ref{figure:Ndet} illustrate our best fits to the detected lines and the hypothetical 3$\sigma$ upper limits when no line is detected.
 
The second-epoch Spitzer spectra of TW~Hya and CS~Cha presented here demonstrate that \neii{} line fluxes are not constant in time and that the infrared continuum emission is also variable. More details on the line and continuum variability of TW~Hya will be presented in an upcoming paper (Najita et al in prep.). The important result for the interpretation of these VISIR spectra is that changes of $\sim$30\% in the line and/or in the continuum are possible even among the class of transition disks.

We do not detect the \neii{} emission line in the Spitzer low-resolution spectrum of HD~34700A. This star is surrounded by a tenuous dust disk and is thus more similar to the optically thin dust disks studied by \citet{pascucci06} rather than to the protoplanetary disks presented in this paper. The \neii{} non-detection suggests that very little or no primordial gas is left in the disk of HD~34700A \citep{pascucci07}, corroborating its identification as a debris disk.  

In the following, we discuss in more detail the 6 protoplanetary disks with spectrally unresolved \neii{} lines in their Spitzer spectra. We separate transition objects from radially continuous dust disks because their VISIR spectra appear remarkably different.

\section{\neii{} emission from the disk of transition objects}\label{Sect:transition}
For the three transition disks TW~Hya, CS~Cha, and T~Cha we detect and spectrally resolve \neii{} lines located near the stellar velocity. In the case of CS~Cha we detect only the emission line but not the continuum.\footnote{The flux calibrated spectrum shown in Fig.~\ref{figure:det} is the mean of the flux calibrated spectra from the 3 datasets, errors at each wavelength are the standard deviations from the 3 spectra on a common wavelength scale}. For all sources \neii{} line fluxes from VISIR spectra are lower than those measured from Spitzer spectra of factors between $\sim$0.5-0.8. One possibility is that the \neii{} emission is extended beyond the 0.4\arcsec{} slit adopted in these observations.  Spatially extended \neii{} emission can certainly arise in outflow sources, as demonstrated by \citet{van09} and by our sample of optically thick dust disks (see Sect.~\ref{Sect:thick}). However, no jets have been reported toward TW~Hya and T~Cha \citep{azevedo07,alcala93} and only a compact (8-10\,mas) outflow in H$\alpha$ has been detected toward CS~Cha \citep{takami03}. In Sect.~\ref{sect:jets} we also show that jets/outflows are not likely to be the main source of \neii{} emission in transition objects. Another explanation for the Spitzer-VLT flux difference is line and/or continuum variability. At least for CS~Cha and TW~Hya we know that line/continuum variability measured from 2-epochs of Spitzer spectra is $\sim$30\%, similar to the \neii{} flux differences from the VLT and Spitzer spectra. We should also keep in mind that at the distance of TW~Hya and T~Cha  (Table~\ref{table:modelparameters}) our slit just covers out to $\sim$10\,AU from the central stars. As discussed later even \neii{} emission from the disk surface can extend beyond several tens of AU.  Further observations with wider slits and different orientation angles are necessary to constrain the spatial extension of the \neii{} emission.

The most important result from the VISIR spectra is that the measured \neii{} lines of transition
disks are relatively narrow (ranging from $\sim$14\,km/s to $\sim$40\,km/s) and peaked near the
stellar velocity (see Table~\ref{table:results}). These line characteristics are consistent with a
disk origin for the emission\footnote{The line width from TW~Hya reported by \citet{herczeg07} is
about 1.6 times larger than our. Their spectrum has a poor S/N and was obtained at an airmass as
large as 1.8.}. We note that TW~Hya, whose disk is almost seen face-on ($4\pm 1^{\circ}$,
\citealt{pontoppidan08}), has the narrower line width.  The inclinations of the disks around CS~Cha and T~Cha are not well constrained but likely closer to edge-on. \citet{espaillat07} successfully model the SED of CS~Cha with a disk viewed at an inclination of 60$^\circ$. T~Cha has a high projectional velocity suggesting that it is viewed almost edge-on \citep{alcala93}.
%is likely closer to edge-on based on its relatively high projectional velocity \citep{alcala93}. 
% \citet{brown07} assume a disk inclination of 75$^\circ$ to model the SED of T~Cha.  
The \neii{} line widths we measure seem to correlate with the disk viewing angle: the narrowest \neii{} line is measured from the almost face-on disk of TW~Hya, a broader line from the inclined disk of CS~Cha, and the broadest line from the almost edge-on disk of T~Cha. This trend also suggests that the \neii{} line profiles are dominated by Keplerian rotation rather than by turbulence.   

\begin{figure}
\includegraphics[scale=0.8]{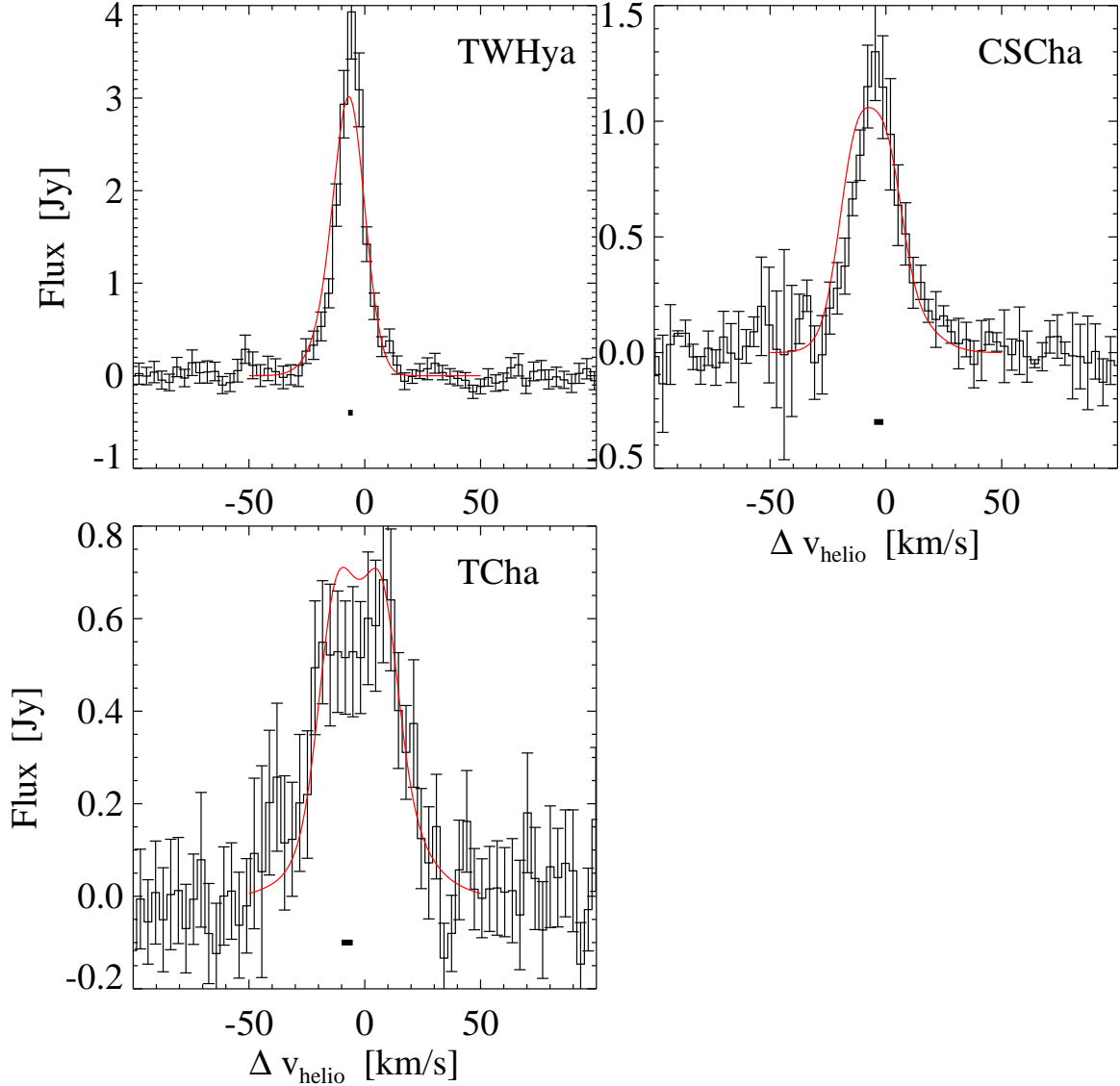}
\caption{Comparison between observed and predicted (red solid line) line profiles for the group of transition disks (velocity in the stellocentric frame). Predicted line profiles are for the standard photoevaporative disk wind assuming an ionizing flux $\Phi$ of $10^{41}$ photons/s and the stellar mass and disk inclination in Table~\ref{table:modelparameters}. Models have been degraded to the spectral resolution of VISIR and scaled by the factors reported in the last column of Table~\ref{table:modelparameters} to match the observed and predicted \neii{} line luminosities. The black thick line below the \neii{} emission shows the total uncertainty in the observed \neii{} peak position. Note the excellent match between the observed and predicted line profile from the almost face-on disk of TW~Hya suggesting that the \neii{} emission originates from a photoevaporative disk wind. The line profiles from CS~Cha and T~Cha are also consistent with model predictions. \label{figure:Alexander}
}
\end{figure}

Further constraints on the region traced by the \neii{} emission can come from modeling line
profiles. In  a recent study, \citet{alexander08} showed that the velocity structure of a
photoevaporative disk wind results in a line profile that is different from that produced by a bound
disk atmosphere. In particular, when a disk is viewed face-on the vertical component of the wind
velocity leads to a profile that is about $\sim$10km/s in width (broader than the thermal line
width) and blueshifted by $\sim$6-7\,km/s. When a disk is viewed closer to edge-on,  line profiles
become broader ($\sim30-40$km/s), are double peaked, dominated by Keplerian rotation, and centered
at the stellar velocity. Fig.~\ref{figure:Alexander} compares the observed \neii{} line profiles
with the line profiles predicted by the standard photoevaporative disk wind model of \citet{font04}
and \citet{alexander08}, the one in which there is no hole in the gas disk. We have chosen this
model because our transition objects are still accreting disk gas, suggesting that there is no hole in the gas disk even if the inner dust disk is largely depleted of small dust grains (Sect.~\ref{Sect:Observations}). The standard photoevaporative disk wind model assumes that the disk atmosphere is heated by the stellar EUV photons at $10^4$\,K and requires just 3 input parameters to compute the \neii{} line profiles: the stellar mass, the disk inclination, and the stellar ionizing flux ($\Phi$). The profiles shown in Fig.~\ref{figure:Alexander} are for the star/disk parameters summarized in Table~\ref{table:modelparameters} and assuming a stellar ionizing flux of $10^{41}$ photons/s.
There is an almost perfect match between the observed and predicted \neii{} profile from the disk of TW~Hya. The almost face-on disk of TW~Hya has a blueshifted \neii{} line fully consistent with the emission originating in a photoevaporative wind moving toward the observer \citep{alexander08}. A static disk atmosphere would result in a line emission centered at the stellar velocity, clearly inconsistent with our observations. Thus, we consider this result as the first strong evidence for a disk being photoevaporated by its central star.

The \neii{} line profiles of CS~Cha and T~Cha are broadly consistent with predictions from the photoevaporative disk wind model (Fig.~\ref{figure:Alexander}), but there are also a few discrepancies, especially for the spectroscopic binary CS~Cha. For this binary the predicted \neii{} line is broader than observed. This could result from a disk viewed at less than 45$^\circ$ inclination, but the emission should be slightly more blue-shifted than observed.

In the simplest assumption that neon atoms are ionized {\it only} by stellar EUV photons, we can use the observed \neii{} line luminosities to estimate the  stellar ionizing flux $\Phi$ reaching the disk surface. The reason is that \neii{} line fluxes scale approximately linearly with $\Phi$ (for $\Phi \le 10^{43}$ photons/s) while the shape of the line profile is insensitive to the value of  $\Phi$ \citep{alexander08}. 
The comparison between the observed and predicted line luminosities yields stellar ionizing fluxes of 2.5, 13, and 1.9$\times 10^{41}$ photons/s for TW~Hya, CS~Cha, and T~Cha respectively. These values have an uncertainty of a factor of 2-3 dominated by the uncertainty in the fraction of neon atoms that exist as Ne$^+$ in the disk surface. \citet{alexander08} keeps this fraction constant to 1/3 while more detailed gas models suggest that it could be as high as 1 \citep{hollenbach09}. The ionizing rates we compute from the \neii{} lines seem plausible for accreting T~Tauri stars. For TW~Hya far-UV observations suggest values of $7\times 10^{39}$ or $5\times10^{41}$ depending on how much EUV emission from the accretion shock is attenuated by the accretion streams \citep{herczegalone07}. As a next step we can use the values of $\Phi$, in combination with star and disk masses, to estimate the disk wind rates and the disk lifetimes if EUV-driven photoevaporation would be the only disk dispersal mechanism. Using the results from \citet{font04} we find disk wind rates ranging from $\sim 2$ to $5 \times 10^{-10}$M$_{\sun}$/yr. At these rates only the relatively low-mass disk of T~Cha (0.003\,M$_\sun$, \citealt{lommen08}) could be photoevaporated in $\sim$10\,Myr. The disks of TW~Hya and CS~Cha would disperse  in several tens of millions of years. However, a more realistic estimate for the disk lifetime should include on-going mass loss onto the star via viscous accretion. In addition, disk lifetimes can be further reduced once the viscous accretion infall rate falls below the photoevaporation rate. At this point a gap opens in the gas disk and the direct EUV flux from the star is expected to disperse the entire disk in only $\sim10^5$\,yr. As we discuss in Sect.~\ref{Sect:discussion}, at least two of our transition objects may be on the verge of opening an inner gap in the gas disk.

Finally, it is worth commenting on the extension of the \neii{} emission in the photoevaporative
disk models. Calculations from  Alexander (priv. comm.) show that more than 90\% of the \neii{}
emission arises inside $\sim 2\times 9 (M_\star/M_\sun)$\,AU. These radial distances are covered by
our 0.4\arcsec{} slit for CS~Cha, are comparable to half the slit width for TW~Hya, and smaller by a
factor of 2 for T~Cha. This shows that we certainly missed some \neii{} flux (especially from the
disks of TW~Hya and T~Cha) even in the case that ionization is solely from EUV photons. Stellar
X-rays could contribute to the gas ionization at larger disk radii \citep{glassgold07} and the X-ray
heated gas might also participate in the photoevaporative flow
\citep{ercolano09}. If X-rays substantially contribute to the \neii{} fluxes then the stellar ionizing fluxes we calculated above should be viewed only as upper limits. It would be extremely interesting in the future to compare the observed \neii{} profiles to those predicted by photoevaporating X-ray-heated gas disks.

\subsection{Can jets explain the \neii{} emission from transition disks?}\label{sect:jets}
The observations of the T~Tau complex from \citet{van09} demonstrate that \neii{} emission can also originate from shocks generated by protostellar outflows/jets. Because of the high ionization potential of neon atoms (21.6\,eV), substantial ionization can be only produced by fast shocks with speeds  $\ge$\,70\,km/s for typical pre-shock densities of 10$^4$\,cm$^{-3}$ (also called J shocks, \citealt{HM89}). For this type of strong radiative shocks the shocked gas moves at almost the shock velocity.  \citet{van09} argue that these high velocities may be reached in the outflow from the T~Tau~S source which has a sight velocity of only $\sim$40\,km/s, but it is likely to be in the plane of the sky.
We now show that it is very unlikely that jets/outflows could explain the
\neii{} emission of transition disks. First, let us consider the case of
TW~Hya. Its disk is seen almost face-on so we would expect any jet/outflow
to be almost in the direction of the observer and the sight velocity to be
close to the actual shock velocity. The sight velocity of the \neii{}
emission from TW~Hya is just -6\,km/s and the line is relatively narrow
(deconvolved FWHM of $\sim$10\,km/s). If this velocity is representative
for the shock velocity as predicted from J shock models, it would be unable
to appreciably ionize neon atoms. Corroborating this assertion,
\citet{hollenbach09} show that the \neii{} line luminosity produced by the
postshock region of a radiative shock depends linearly on:
$\dot{M}_w\,v_s^2$, where $\dot{M}_w$ is the protostellar mass wind loss
rate and $v_s$ is the shock velocity. The wind mass loss rate scales with
the stellar mass accretion rate as $\dot{M}_w \simeq 0.01-0.1\times
\dot{M}_{acc}$ \citep{hartigan95,WH04}. Given the low stellar accretion
rates of transition disks, the \neii{} fluxes reported in
Table~\ref{table:results} could be reached only with shock velocities
ranging from several hundreds of km/s to thousands of km/s, clearly
inconsistent with the small velocity shifts and relatively small FWHM of
the observed \neii{} lines. Finally, line widths from shocked gas are
expected to be quite broad (several tens of km/s) regardless of the
observer viewing angle (e.g., \citealt{hartigan87}), thus shocked gas
cannot reproduce the observed trend between the FWHM of \neii{} lines and
the disk inclination. Although shocks generated by jets/outflows are
unlikely to explain the \neii{} emission from transition disks, they may
dominate the \neii{} emission from less-evolved optically thick dust disks
(see \citealt{guedel09} and next Section).

\section{Jets/outflows dominate the \neii{} emission of  classical optically thick disks}\label{Sect:thick}
The three optically thick dust disks observed here have spectrally unresolved \neii{} fluxes comparable to those of transition disks. Nevertheless, we do not detect any \neii{} line close to the stellar velocity in their VISIR spectra. Only in the case of Sz~73 we detect  a broad ($\sim$60\,km/s) emission line at $\sim$-100\,km/s  with respect to the stellar velocity and a line flux that is just $\sim$30\% lower than the Spitzer \neii{} flux. The large FWHM in combination with the large blueshift point to a jet/outflow origin for the \neii{} emission. The case of Sz~73 well illustrates that when jets/outflows are present the \neii{} emission they produce can dominate over the disk emission. This is in line with the observations of the T Tau complex  from \citet{van09}.
  
The \neii{} non-detections of Sz~102 and VW~Cha are also consistent with this scenario.  If the \neii{} emission measured with Spitzer would originate in a disk, only relatively broad lines (FWHM $>$50km/s), compatible only with close to edge-on disks, would be undetected in our VISIR spectra. Interestingly, VW~Cha and Sz~102 are known to have outflows at position angles (from N to E) of 90 and 99 degrees respectively \citep{bally06,krautter86,gh88}. Since we did not change the default slit orientation of VISIR (PA=0$^\circ$) we acquired spectra in a direction perpendicular to the known outflows. Thus, our non-detections can be reconciled with the Spitzer observations if most of the \neii{} emission arises in jets/outflows, which can be verified with future high-resolution spectra covering the region of the known outflows.

%Though the sample statistics is  still rather limited, it is tempting to conclude that most of the \neii{} emission from optically thick (non-transition) disks does not originate in a disk but rather in jets/outflows.  

\section{On the onset of centrally-driven photoevaporation}\label{Sect:discussion}
 
Viscous accretion is thought to be the dominant disk dispersal mechanism but alone cannot explain the 
relatively short ($\lesssim 10$\,Myr) disk dispersal timescales inferred from observations (e.g. review by \citealt{dullemond07}).
High-energy stellar photons can heat and ionize the disk surface and induce
photoevaporation (i.e. disk mass loss) from outer disk regions. Exactly
when photoevaporation becomes an important disk dispersal mechanism and how
much disk mass it can remove is still a matter of debate. FUV and X-ray
photons can penetrate much larger column densities of gas than stellar EUV
photons. As a consequence, they are expected to drive efficient
photoevaporation from the early times of disk evolution at rates that may
exceed the EUV photoevaporation rate by one or even two orders of
magnitudes \citep{gorti09,ercolano09}. Stellar EUV photons start to penetrate the disk wind, and thus reach the disk surface, only when accretion rates fall below $\sim 10^{-8}$M$_{\sun}$/yr, corresponding to hydrogen column density screens of N(H)$\lesssim$10$^{17}$\,cm$^{-2}$  \citep{hollenbach09}. Although EUV photons likely produce little mass loss at all stages \citep{gorti09}, they can help to quickly ($\sim 10^5$\,yr) disperse remaining disk gas once stellar accretion rates fall below a few $10^{-10}$M$_{\sun}$/yr \citep{alexander06a}. 

Recently, \citet{currie09} have argued that the large number of evolved {\it dust} disks in the 5\,Myr-old cluster NGC~2362 is inconsistent with predictions from the standard UV photoevaporation model, in which  EUV-driven photoevaporation follows viscous accretion \citep{clarke01,alexander06b}. Indeed, these models predict a paucity of evolved disks because once the accretion inflow rates fall below the photoevaporation rate the disk  dispersal is fast ($\sim 10^5$\,yr). However, model predictions are directly relevant only to the evolution of the gas disk component. In order to translate the gas evolution into dust evolution, \citet{alexander06b} made the simplistic assumption that the dust is coupled to the gas throughout the entire disk evolution. This is certainly unrealistic because grain growth, observed in almost all protoplanetary disks (e.g. \citealt{natta07}), results in dust settling, i.e. decoupling of the dust from the gas disk \citep{dullemond04}. A more direct way to test photoevaporation models is to constrain the evolution of the gas disk through observations of gas lines tracing the accreting as well as the possibly photoevaporating material. In what follows we will take this approach to speculate on the onset of photoevaporation.

We start by discussing our sample of probably less evolved classical optically thick disks with radially continuous dust distribution.
The H$\alpha$ equivalent widths from these three systems are greater than 50\,\AA{} \citep{hughes94,guenther07}, indicating that the central stars are accreting nebular gas. This is corroborated by the detection of outflows and large near-infrared excess emission, both characteristic to actively accreting disks  \citep{hartigan95}. In these systems we do not find evidence for on-going disk photoevaporation. The \neii{} line flux upper limits at the stellar velocity imply that not only EUV photons but also stellar X-rays are efficiently screened by the protostellar wind and accretion columns, requiring large gas column densities (N$_H \ga 10^{22}$\,cm$^{-2}$, e.g. \citealt{guedel07}). If column densities exceed $\sim 10^{24}$\,cm$^{-2}$, FUV photons are also significantly attenuated \citep{hollenbach09}. For lower column densities,  photoevaporation driven by FUV photons could be occurring and we would not detect it because FUV photons have energies lower than the ionization potential of neon atoms (21.6\,eV). These systems may be at the same evolutionary stage as the T Tau complex.

Recently, \citet{najita09} reported spectrally resolved \neii{} lines from two disks in the Taurus-Auriga star-forming region: AA~Tau, a classical optically thick disk whose star presents periodic accretion rate bursts \citep{bouvier07}; and GM~Aur, a transition disk whose central star is accreting at a level comparable to that of classical T Tauri stars (10$^{-8}$M$_{\sun}$/yr, \citealt{gullbring98}). The \neii{} lines are found to be centered at the stellar velocity suggesting that the emission originates from a static disk atmosphere very likely heated by stellar X-rays. However, the measured fluxes are factors of 2 and 3 lower than the  spectrally unresolved fluxes from Spitzer spectra. One possible explanation is that most of the undetected \neii{} emission originates far from the central stars, in ambient gas shocked by a jet/outflow. This seems very plausible for AA~Tau, whose optical spectrum has forbidden emission lines typical of high-velocity jets (e.g. [NII] 6530\,\AA , \citealt{hirth97}) and a jet has also been imaged with HST/STIS and Goddard Fabry Perot spectrograph \citep{cox05}. The \neii{} emission reported by \citet{najita09} does not seem to trace photoevaporating gas in these two systems, possibly because X-ray heated gas producing \neii{} emission is slightly cooler ($\sim$5,000\,K) and thus more bound than the EUV-heated gas. 

Finally, the transition disks presented here have \neii{} line profiles consistent with emission from a photoevaporative disk wind indicating on-going photoevaporation at this evolutionary stage. In the case of TW~Hya, the stellar accretion rate is close to the limit where EUV-driven photoevaporation could start opening a gap in the gas disk. The small but varying H$\alpha$ equivalent width of T~Cha suggests that this transition object may also be on the verge of clearing a gas hole.  Interestingly, TW~Hya and T~Cha are also thought to be relatively old, older than 5\,Myr (\citealt{webb99,ancker98}, see also Table~\ref{table:prop}). CS~Cha is the youngest among our transition disks. Its upper limit to the stellar accretion rate is close to what is needed for EUV photons to start penetrating the disk wind, placing it at an earlier evolutionary stage than TW~Hya and T~Cha (though the evolution of the inner disk is certainly affected by presence of the stellar companion). 

In overall, these results support the evolutionary picture in which EUV-driven photoevaporation occurs at later times in the disk evolution and the mass loss rates can be only modest. We should also note that if stellar X-rays substantially contribute to the \neii{} emission of transition disks then the stellar ionizing flux $\Phi$ calculated in Sect.~\ref{Sect:transition} would be only upper limits, further reducing the efficiency of EUV-driven photoevaporation. Future observations should aim at expanding the sample of classical and transition disks with spectrally resolved \neii{} lines and accretion rate measurements. In addition, it would be extremely valuable to compute \neii{} line profiles from X-ray heated photoevaporating gas that could be directly compared to observations.

%Though the sample statistics is still rather limited, it is tempting to conclude that efficient photoevaporation starts when stellar accretion rates drops below  $\sim 10^{-8}$ M$_{\sun}$/yr.

%% The equation environment wil produce a numbered display equation.

\section{Summary}
We used the high-resolution spectrograph VISIR on the VLT to spectrally resolve \neii{} emission lines
previously detected with the Spitzer Space Telescope in a sample of young protoplanetary disks. Our main results can be summarized as follows:
\begin{itemize}
\item We detect and spectrally resolve the \neii{} lines from the 3 transition disks in our sample.
The line centroids, widths, and intensities indicate that the \neii{} emission arises in a disk. Models of disk wind photoevaporation well reproduce the observed line profiles strongly suggesting  on-going photoevaporation driven by stellar EUV photons at this evolutionary stage.
\item From the sample of optically thick radially continuous dust disks we only detect a strong and broad \neii{} emission blueshifted by -100\,km/s toward Sz~73. This points to a jet/outflow origin for the \neii{} emission. The \neii{} non-detection in the other two systems with known jets also suggest that most of the spectrally unresolved \neii{} emission detected with Spitzer does not arise from the disk but rather from surrounding gas shocked by jets/outflows.
\end{itemize}
These results provide the first observational evidence for centrally-driven disk photoevaporation. Because transition disks have already undertaken significant evolution in comparison to the radially continuous dust disks, our results demonstrate that EUV-driven photoevaporation can occur only at a late stage in the disk evolution.

%% The \notetoeditor{TEXT} command allows the author to communicate
%% information to the copy editor.  This information will appear as a
%% footnote on the printed copy for the manuscript style file.  Nothing will
%% appear on the printed copy if the preprint or
%% preprint2 style files are used.

%% The eqnarray environment produces multi-line display math. The end of
%% each line is marked with a \\. Lines will be numbered unless the \\
%% is preceded by a \nonumber command.
%% Alignment points are marked by ampersands (&). There should be two
%% ampersands (&) per line.

%% If you wish to include an acknowledgments section in your paper,
%% separate it off from the body of the text using the \acknowledgments
%% command.

%% Included in this acknowledgments section are examples of the
%% AASTeX hypertext markup commands. Use \url without the optional [HREF]
%% argument when you want to print the url directly in the text. Otherwise,
%% use either \url or \anchor, with the HREF as the first argument and the
%% text to be printed in the second.

\acknowledgments
We thank R. D. Alexander for proving the \neii{} line profiles from photoevaporative disk winds 
and for extremely useful discussions. We would also like to thank F. Lahuis for making available 
the high-resolution Spitzer/IRS spectra of Sz~73, Sz~102, T Cha, and VW~Cha and L. Decin for 
the MARCS model atmosphere of HR~6705. We are also grateful to D. Hollenbach, U. Gorti, 
B. Ercolano, J. Najita, J. Carr, and D. Apai 
for valuable discussions, and an anonymous referee for helpful suggestions. 
IP is pleased to acknowledge support through NASA contract 90035375.

%% To help institutions obtain information on the effectiveness of their
%% telescopes, the AAS Journals has created a group of keywords for telescope
%% facilities. A common set of keywords will make these types of searches
%% significantly easier and more accurate. In addition, they will also be
%% useful in linking papers together which utilize the same telescopes
%% within the framework of the National Virtual Observatory.
%% See the AASTeX Web site at http://www.journals.uchicago.edu/AAS/AASTeX
%% for information on obtaining the facility keywords.

%% After the acknowledgments section, use the following syntax and the
%% \facility{} macro to list the keywords of facilities used in the research
%% for the paper.  Each keyword will be checked against the master list during
%% copy editing.  Individual instruments or configurations can be provided 
%% in parentheses, after the keyword, but they will not be verified.

{\it Facilities:} \facility{VLT (VISIR)}, \facility{Spitzer (IRS)}.

\clearpage

%% Use the figure environment and \plotone or \plottwo to include
%% figures and captions in your electronic submission.
%% To embed the sample graphics in
%% the file, uncomment the \plotone, \plottwo, and
%% \includegraphics commands
%%
%% If you need a layout that cannot be achieved with \plotone or
%% \plottwo, you can invoke the graphicx package directly with the
%% \includegraphics command or use \plotfiddle. For more information,
%% please see the tutorial on "Using Electronic Art with AASTeX" in the
%% documentation section at the AASTeX Web site,
%% http://www.journals.uchicago.edu/AAS/AASTeX.
%%
%% The examples below also include sample markup for submission of
%% supplemental electronic materials. As always, be sure to check
%% the instructions to authors for the journal you are submitting to
%% for specific submissions guidelines as they vary from
%% journal to journal.

%% This example uses \plotone to include an EPS file scaled to
%% 80% of its natural size with \epsscale. Its caption
%% has been written to indicate that additional figure parts will be
%% available in the electronic journal.

%\begin{figure}
%\epsscale{.80}
%\plotone{f1.eps}
%\caption{Derived spectra for 3C138 \citep[see][]{heiles03}. Plots for all sources are available
%in the electronic edition of {\it The Astrophysical Journal}.\label{fig1}}
%\end{figure}

\clearpage

%% Here we use \plottwo to present two versions of the same figure,
%% one in black and white for print the other in RGB color
%% for online presentation. Note that the caption indicates
%% that a color version of the figure will be available online.
%%

%\begin{figure}
%\plottwo{f2.eps}{f2_color.eps}
%\caption{A panel taken from Figure 2 of \citet{rudnick03}. 
%See the electronic edition of the Journal for a color version 
%of this figure.\label{fig2}}
%\end{figure}

%% This figure uses \includegraphics to scale and rotate the still frame
%% for an mpeg animation.

\clearpage

\begin{deluxetable}{lcc c ccc}
\tabletypesize{\scriptsize}
%\rotate
\tablecaption{Stellar properties and \neii{} fluxes measured with the Spitzer Space Telescope \label{table:prop}}
\tablewidth{0pt}
\tablehead{
\colhead{Source} & \colhead{2MASS J\tablenotemark{a}} & \colhead{v$_{\rm rad}$\tablenotemark{b}}&\colhead{Age} &\colhead{Ref.} & \colhead{[NeII] flux} &\colhead{Ref.}\\
\colhead{} & \colhead{}                               &  \colhead{(km/s)} &  \colhead{(Myr)} &  \colhead{}  &\colhead{($\times 10^{-14}$\,\uflux)}& \colhead{}  
}
\startdata
HD~34700Aa\tablenotemark{*}& 05194387+0539406&68.0$\pm$0.2 & 1-10&1    & -- & 11 \\
%Ab has vr = -27.98 km/s
TW~Hya  & 11015191-3442170& 12.2$\pm$0.5   & $\sim$10  &2,3   & 4.5$\pm$0.3,5.7 &  11,12 \\
CS~Cha\tablenotemark{*}  & 11022491-7733357& 14.9$\pm$2     & $\sim$2& 4,5,6 & 3.4$\pm$0.1,4.3 & 11,13 \\
VW~Cha\tablenotemark{*}  & 11080148-7742288& 17.2$\pm$0.5   & $\sim$2& 5,6   &2.8$\pm$0.4&  14\\
T~Cha   & 11571348-7921313& 14.0$\pm$1.3   & $>$12 &5,7   &3.2$\pm$0.2& 14\\
Sz~73   & 15475693-3514346& -3.3$\pm$2.5   & 2.6-5.4& 8,9    &1.6$\pm$0.2&  14\\
Sz~102  & 16082972-3903110& 5 & ...&10,9   &  3.6$\pm$0.1&  14\\
\enddata
\tablenotetext{a}{The 2MASS source name includes the J2000 sexagesimal, equatorial position in the form: hhmmssss+ddmmsss 
\citep{cutri03}.}
\tablenotetext{b}{Stellar heliocentric radial velocity}
\tablenotetext{*}{These stars are part of multiple stellar systems. HD~34700A is a spectroscopic binary and has two visual companions \citep{sterzik05}.
CS~Cha was recently found to be a long-period spectroscopic binary \citep{guenther07}. VW Cha has a visual companion at $\sim$0.7\arcsec{} \citep{brandner96}.}
\tablerefs{
(1) \citealt{sterzik05}; (2) \citealt{weintraub00};
(3) \citealt{barrado06};
(4) \citealt{jg01}; (5) \citealt{guenther07};
(6) \citealt{luhman08}; (7) \citealt{ancker98};
(8) \citealt{melo03}; (9) \citealt{hughes94};
(10) \citealt{gh88}; (11) this work, line fluxes with errorbars;
(12)  \citealt{ratzka07}, second entry; (13) \citealt{espaillat07}, second entry;
(14) \citealt{lahuis07} }
\end{deluxetable}
%% Notes on spTy
%% HD34700 - G0 (Moira et al. 2001); TWHya - K7 (Webb et al. 1999) ; CSCha Guenther et al. 2007; VWCha K2-K7 Lahuis et al. 2007; T Cha - G8 Guetnther et al. 2007;
%% Sz 73  M0 Hughes et al. 1994; Sz 102 K0 Hughes et al. 2007

\begin{deluxetable}{lccccclcc}
\tabletypesize{\scriptsize}
%\rotate
\tablecaption{Summary of the observations. The observing time (in U.T.), the airmass, 
and the heliocentric radial velocity corrections (v$_{\rm helio}$) are given at the beginning and at the
end of the observations. \label{table:log}}
\tablewidth{0pt}
\tablehead{
\colhead{Source} & \colhead{Date} & \colhead{U.T.}& \colhead{t$_{exp}$} & \colhead{Airmass} &\colhead{v$_{\rm helio}$}& \colhead{Calibrator} & \colhead{t$_{exp}$}& \colhead{Airmass} \\
\colhead{} & \colhead{yyyy-mm-dd} &\colhead{hh:mm}&\colhead{(s)} & \colhead{} &\colhead{(km/s)} &\colhead{}  &\colhead{(s)} & \colhead{}
}
\startdata
TW~Hya & 2008-02-20 &02:37/04:07&3600 & 1.3/1.1 & 13.16/13.03&HD~90957(K5III) & 240 & 1.0(F) \\
CS~Cha & 2008-02-20 &05:00/06:29&3600 & 1.7/1.7 & 12.01/11.97&BV441(M0III) & & \\
VW~Cha & 2008-02-20 &07:07/07:15&3480 & 1.7/1.9 & 12.06/12.03&HD~92682(K3II) & 240 & 1.6(P) \\
VW~Cha & 2008-02-20 &07:35/09:01&3480 & 1.7/1.9 & 12.06/12.03&HD~105340(K2II) & 240 & 1.7(F) \\
%VW~Cha & 2008-02-20 &09:39/09:56&720  & 1.9/2.0 & 12.03/12.03&HD~105340 &  240 & 1.7(P) \\
\hline
HD~34700A& 2008-02-21&01:22/02:51&3600 & 1.2/1.6 & -27.40/-27.55&EPS~Tau(G9.5III) & 240 & 1.6(P) \\
T~Cha   & 2008-02-21&03:34/06:37&7200 & 1.9/1.7 & 12.99/12.94&BV441(M0III)&240 & 1.8(P) \\
        &            & &     &         & &HD~105340(K2II)&360 & 1.6(F) \\
Sz~102  & 2008-02-21&07:41/09:11&3600 & 1.3/1.1 & 28.74/28.64&HD~136422(K5III)&360 & 1.2(P) \\
        &            & &     &          & &HD~139127&600 & 1.1(F) \\
\hline
CS~Cha  & 2008-02-22&02:23/04:08&4140 & 1.9/1.7 & 12.03/12.0&BV441(M0III) & 240 & 1.9(P) \\
CS~Cha  & 2008-02-22&04:28/06:00&3600 & 1.7/1.7 & 11.99/11.95&HD~92682(K3II) & 360& 1.6(F) \\
Sz~73   & 2008-02-22&06:58/09:17&5400 & 1.4/1.0 & 29.36/29.19&HD~136422(K5III) & 360 & 1.4(P) \\
        &            & &     &         & &HD~139127(K4.5III) & 600 & 1.1(F) \\
\enddata
%\tablenotetext{a}{Heliocentric radial velocity correction at the time of the observation}
\end{deluxetable}

%\begin{deluxetable}{lcccc}
%\tabletypesize{\scriptsize}
%\rotate
%\tablecaption{Properties of the standard stars for myself
%\label{table:stdprop}}
%\tablewidth{0pt}
%\tablehead{
%\colhead{Std} & \colhead{SpTy} & \colhead{Model Flux at 12.81um} &\colhead{Exposure Time}& \colhead{S/N} \\
%\colhead{} & \colhead{} &\colhead{(Jy)} & \colhead{(min)} & \colhead{at  12.808\micron}
%}
%\startdata
%HD~90957& K5III& 3.2 & 240& 10 \\
%BV411   & M0III& 21.5&240 & 35 \\
%HD~92682& K3II& 4.6 & 240& 7\\
%HD~105340& K2II& 3.6 & 240& 8\\
%\hline
%EPS~Tau  &G9.5III& 3.6 &240& 12 \\
%HD~105340& K2II& 3.6 & 360& 10\\
%HD~136422&K5III&  29.5&  360& 40  \\
%HD~139127&K4.5III&  10.5& 600 & 30 \\
%\hline
%BV441   & M0III &  21.5& 240 & 34\\
%HD~92682& K3II&  4.6& 360& 13\\
%HD~136422&K5III& 29.5 &  360& 53 \\
%HD~139127&K4.5III& 10.5  & 600 & 52\\
%\enddata
%\end{deluxetable}

\begin{deluxetable}{lcccc}
%\tabletypesize{\scriptsize}
%\rotate
\tablecaption{Summary of the VISIR results. For \neii{} detections we assume a Gaussian for the line profile
and a first order polynomial for the continuum. The last column gives the velocity of the  peak emission in
the stellocentric frame. 
\label{table:results}}
\tablewidth{0pt}
\tablehead{
\colhead{Source}  &\colhead{\neii} & \colhead{Line Flux\tablenotemark{*}} & \colhead{FWHM} & \colhead{v$_{\rm peak}$} \\
\colhead{} & \colhead{detection?} &\colhead{($\times 10^{-14}$\,\uflux)} & \colhead{(km/s)} & \colhead{(km/s)}
}
\startdata
HD~34700A& N& $<0.5$  & & \\
TW~Hya  &  Y& 3.8$\pm$0.3 & 14.6$\pm$0.7 & -6.2$\pm$0.3 \\
CS~Cha  &   Y\tablenotemark{a}& 2.3$\pm$0.2  & 27$\pm$2&-3.3$\pm$0.7 \\
VW~Cha  &  N& $<0.6$ & & \\
T~Cha   &  Y&  2.2$\pm$0.3  & 42$\pm$4 & -4$\pm$2 \\
Sz~73   &  Y& 1.1$\pm$0.2 & 60$\pm$8  & -99$\pm$3 \\
Sz~102  &  N&  $<0.5$ &  & \\
\enddata
\tablenotetext{*}{In the case of non-detections we provide 3$\sigma${} upper limits to the line flux assuming 
a width of 10\,km/s, equal to the instrument resolution. If the line is broader the flux upper limit would increase
proportionally with the line FWHM by the factor (FWHM/10). }
\tablenotetext{a}{In this case the continuum emission is not detected}
%\tablenotetext{a}{This is a 3.5$\sigma$ detection}
\end{deluxetable}

\begin{deluxetable}{lccccc}
%\tabletypesize{\scriptsize}
%\rotate
\tablecaption{Stellar and disk parameters used to generate the \neii{} profiles in Fig.~\ref{figure:Alexander}.
Stellar ionizing flux $\Phi$ were fixed to $10^{41}$ photons/s. The last column provides the ratio between the observed and predicted \neii{} line luminosity, which is also the scaling factor for the ionizing flux $\Phi$. The comparison between models and observations yields $\Phi = $ 2.5, 13, and 1.9$\times 10^{41}$ photons/s for TW~Hya, CS~Cha, and T~Cha respectively with an unceratinty of a factor of 2-3.  
\label{table:modelparameters}}
\tablewidth{0pt}
\tablehead{
\colhead{Source} & \colhead{M$_\star$} &\colhead{Disk inclination} & \colhead{Distance}&\colhead{Ref.} & \colhead{L$_{\rm obs}$/L$_{\rm mod}$}  \\
\colhead{} &\colhead{(M$_\sun$)} &\colhead{($^\circ$)} &  \colhead{(pc)} & \colhead{} & \colhead{}
}
\startdata
TW~Hya  & 0.7 & 4  & 51  &1,2,3 & 2.5  \\
CS~Cha  & 0.9 & 45\tablenotemark{*} & 160 &4,5 & 13    \\
T~Cha   & 1.5 & 75 & 66  &6,7 & 1.9 \\
\enddata
\tablenotetext{*}{Chosen to provide a good match to the \neii{} line centroid and FWHM}
\tablerefs{
(1) \citealt{muzerolle00}; 
(2) \citealt{pontoppidan08};
(3) \citealt{mamajek05}; 
(4) \citealt{espaillat07};
(5) \citealt{luhman08};
(6) \citealt{alcala93};
(7) \citealt{ancker98}
}
\end{deluxetable}

\end{document}